%% file: main.tex
\documentclass[aps,prl,twocolumn,groupedaddress, nobibnotes]{revtex4-2}

\usepackage{float,hyperref}
\usepackage{graphicx,color}
\usepackage{url}
\usepackage{adjustbox}
\usepackage{amsmath}
\usepackage{amssymb,bm} 
\usepackage{physics} 
\usepackage[dvipsnames]{xcolor}
\usepackage{multirow}
\hypersetup{colorlinks=true,urlcolor=blue}

\newcommand{\angstrom}{\mbox{\normalfont\AA}}
\newcommand{\cT}{{\cal T}}
\newcommand{\cI}{{\cal I}}
\newcommand{\cP}{{\cal P}}

\begin{document}

\title{Light-element and purely charge-based topological materials}

%Author Information
\author{Nassim Derriche}
\email[Corresponding author: ]{nderriche@phas.ubc.ca}
\author{Marcel Franz}
\email[]{franz@phas.ubc.ca}
\author{George Sawatzky}
\email[]{sawatzky@physics.ubc.ca}
\affiliation{Department of Physics and Astronomy \& Stewart Blusson Quantum Matter Institute,
University of British Columbia, Vancouver BC, Canada V6T 1Z4}

%\date{\today}

\begin{abstract}
We examine a class of Hamiltonians characterized by interatomic, interorbital even-odd parity hybridization as a model for a family of topological insulators without the need for spin-orbit coupling. Non-trivial properties of these materials are exemplified by studying the topologically-protected edge states of $s$-$p$ hybridized alkali and alkaline earth atoms in one and two-dimensional lattices. In 1D the topological features are analogous to the canonical Su–Schrieffer–Heeger model but, remarkably, occur in the absence of dimerization. Alkaline earth chains, with Be standing out due to its gap size and near particle-hole symmetry, are of particular experimental interest since their Fermi energy without doping lies directly at the level of topological edge states. Similar physics is demonstrated to occur in a 2D honeycomb lattice system of $s$-$p$ bonded atoms, where dispersive edge states emerge. Lighter elements are predicted using this model to host topological states in contrast to spin-orbit coupling-induced band inversion favoring heavier atoms.
\end{abstract}

\maketitle

\section{Introduction}

The study and search for topologically insulating materials is an important part of contemporary materials science research owing to the unique and often useful properties such systems exhibit. Topological insulators, by definition, host states physically localized on the surfaces or edges of a sample that energetically lie in a gap of the bulk material’s band structure \cite{3d_topological_material_original, 3d_topological_material_followup, quantum_spin_hall_effect_original, quantum_spin_hall_effect_realisation, colloq_topo, topologial_insulators_review, topological_textbook}. Because the emergence of these surface states is rooted in topology and the associated symmetries of the Hamiltonian, their existence can be robustly protected from impurities or external perturbations. This makes topological insulators appealing platforms for various applications such as the low-dissipation semiconductor devices and, due to the spin-momentum locking of surface states, also for magnetic applications \cite{review_TI,magnetic_application,mosfet_ti_1,mosfet_ti_2}. A plethora of systems has been shown to be topological insulators but the search for new candidates is ongoing, motivated by the need for specific physical properties and materials suitable for technological applications.

We examine and analyze here a class of Hamiltonians that host topological phases, applicable to materials in which interorbital hybridization of even (\textit{gerade}) and odd (\textit{ungerade}) atomic orbitals is a significant part of the low energy electronic physics. As prototypical examples, we focus on systems dominated by nearest-neighbor sigma hybridization between $s$ and $p$ orbitals; due to their electron configuration, alkali and alkaline earth atoms constitute excellent systems to which our model can be applied. We concentrate on lower-dimensional materials since, compared to 3D systems, they more commonly feature band gaps extending over the full Brillouin Zone. These systems however have different electronic structures and Fermi energies than their 3D counterparts. The fundamental physics of interest for such materials is at first most clearly illustrated in a one-dimensional chain; such a system has been studied in the framework of the modern theory of polarization \cite{modern_theory_of_polarization_precursor, modern_theory_of_polarization_original, modern_theory_of_polarization_summary, jeroen_thesis, jeroen_screening_organic, polarization_spaldin}, but the aim of the 1D-focused part of this article is to formally describe the topology involved by analyzing a simple Hamiltonian containing the key even-odd orbital hybridization (similarly to the Shockley model) \cite{shockley_original, sp_1d_shockley} in order to show that it is enough for topological phases to emerge by using realistic parameters calculated with an \textit{ab initio} approach, specifically density functional theory (DFT) in this work \cite{lif_pol, sp_pol_original}. 

For materials to exhibit this type of topological behaviour, their band structure needs to feature a gap induced by the avoided crossing of an even and an odd parity band. For this to be of practical importance, their chemical potential should lie in this gap either naturally or by shifting it accordingly through small electron or hole doping. This makes the topologically-protected edge states emerging in the gap experimentally accessible. The alkali and alkaline earth linear chains and 2D honeycomb lattices discussed in detail later feature such gaps. As shown in Table \ref{table:phase_table}, while the Fermi energy in the conducting alkali chains is separated from the gap, for alkaline earth chains which have two valence electrons per atom it lies in the gap and, additionally, directly at the energy of the degenerate edge states $E_{Edge}$ in the topological phase \cite{be_chain_dft_electronic_structure}. This is because the two degenerate gap states appearing in a topological finite chain are "taken" from the bulk bands (one from the lower and one from the upper one). Since the total number of available electronic states does not change, a fully filled lower band in a trivial phase will lead to one of the two gap states being occupied after a topological transition.  Consequently, we present data focused more on the alkaline earth materials in this work. Of course, there are likely alloys and intermetallic compounds in which these elements play a dominant role in the valence electronic structure where this also could occur, highlighting the rather broad classes of materials that potentially exhibit this type of topological behavior. Naturally, the discussion of a topological insulator involving $p$ orbitals brings graphene to mind as a historically important system \cite{graphene_original, semenoff_graphene_tight_binding, graphene_review}. The model ingredients we consider here are however fundamentally different and spinless; proposals for graphene-based \cite{graphene_1, graphene_2} and even 2D trigonal $s$-$p$ topological insulators \cite{trigonal} are characterized by the rise of the quantum spin Hall effect that requires strong spin-orbit coupling, whose strength is increased through the adsorption of heavy atoms or a substantial external field \cite{adsorb}.

\begin{figure}
\includegraphics[width=\columnwidth]{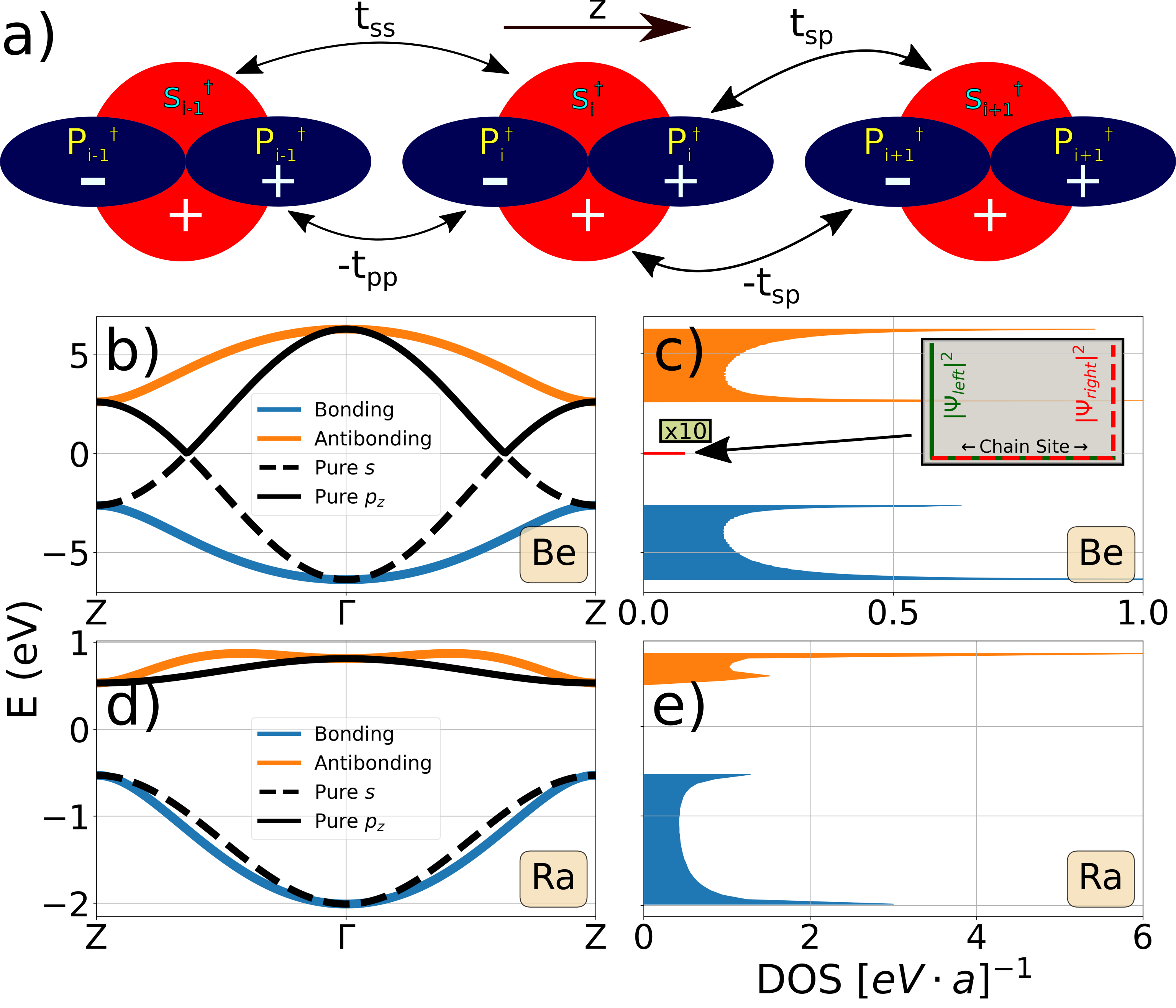}
\caption{$s$-$p$ chain results. a) Diagram of a uniform chain of $s$ and $p$ orbitals and the allowed nearest-neighbor hopping processes. b) Band structure calculated with Eq. \eqref{eq:hamiltonian} (black curves have $t_{sp}$ = 0 set). The high symmetry points are $\Gamma$ ($k= 0$) and $Z$ ($k = \pm \frac{\pi}{2a}$). c) Density of states of a $N = 5000$ chain of Be atoms, where the value for the edge states in red has been multiplied by 10 for visual clarity. Inset: charge density of the two degenerate edge states in the gap. d) and e) are the same as b) and c) for Ra chains which don't exhibit gap states. The Fermi energy is set at $E=0$, and the \textit{ab initio} parameters used are listed in Table \ref{table:phase_table}.}\label{fig:Chain_Be_Ra_Results}
\end{figure}

\section{Results and Discussion}

We consider a one-dimensional infinite chain of equidistant atoms that each host an $s$ orbital and a $p$ orbital with lobes aligned along the chain axis, as illustrated in Fig. \ref{fig:Chain_Be_Ra_Results} (a). Such chains are known to not be susceptible to a longitudinal Peierls transition to a dimerized system contrary to a H or C chain \cite{derriche_peierls, dft_na}, but it must be noted that transversal structural instabilities may exist \cite{mazin_na_2008}; we however constrain ourselves to one-dimensional uniform chains (and later two-dimensional hexagonal structures) for the purpose of highlighting the topological properties of these systems. The tight binding Hamiltonian in reciprocal space in an ordered $(s,p)$ basis reads 
\begin{equation}\label{eq:hamiltonian}
h(k) = \mqty(-2t_{ss}\cos(ak) - \frac{\Delta}{2} & -2it_{sp}\sin(ak) \\ 2it_{sp}\sin(ak) & 2t_{pp}\cos(ak) + \frac{\Delta}{2}),
\end{equation}
where $\Delta$ is the on-site energy difference between the $p$ and $s$ orbitals in the chain, $a$ is the lattice constant and $k$ represents the crystal momentum in the chain direction. In Fig. \ref{fig:Chain_Be_Ra_Results} (b) and (d), the band structure using beryllium and radium parameters listed in Table \ref{table:phase_table} respectively are displayed. Comparing the bands with ($t_{sp}>0$) and without ($t_{sp}=0$) $s$-$p$ mixing in Fig. \ref{fig:Chain_Be_Ra_Results} shows that the light Be chain bands in (b) anticross, leading to the manifestation of two degenerate edge states in the gap in (c), while the heavy Ra chain in (d) has no crossing at all and thus no topological states in (e). We note that the $p_x$ and $p_y$ orbitals, which are higher in energy from the chain-aligned $p_z$ orbitals by crystal field effects, are not important to this model \cite{be_chain_dft_electronic_structure}. For both the alkali and alkaline earth chains which respectively have one and two valence electrons per atom, the $s$-$p$ $\sigma$-bonding describes the active physics while the hybridized orbitals formed from $\pi$-bonding between the $p$ orbitals perpendicular to the chain direction are unoccupied without doping and do not hybridize with the occupied states. The two alkaline earth extremes were selected to highlight and contrast the mass dependence of the $s$-$p$ hybridization topological phenomenon with the more common spin-orbit coupling induced topology. As shown in the gap data found in Table \ref{table:phase_table}, the lighter elements host topological phases and the band gap monotonically decreases for alkali and alkaline earth materials respectively until the gap closes, indicating a transition from the topological to the trivial phase in which no avoided crossing occurs, and then the gap increases monotonically. On the other hand, the strength of relativistic effects, including spin-orbit coupling, increases strongly with higher nuclear charge and principal quantum number $n$ of the valence band states \cite{quantum_electrodynamics_book_lifshitz, relativity_atoms, rel_perspective}. This, along with the decrease of the bandwidths with nuclear charge controlled by $t_{ss}$ and $t_{pp}$ due to larger equilibrium lattice constants $a$ and the increasing emergence of nodes in the orbital wavefunctions affecting orbital overlaps, explains why heavier elements exhibit increased separation between $ns$ and $np$ bands thus avoiding band crossing in the first place \cite{harrison_book}.

\begin{table}[h!] 
\centering
\begin{tabular} {|c|c|c|c|c|c|c|c|c|c|} 
\hline
\multirow{2}{*}{Element} & \multirow{2}{*}{$a$} & \multirow{2}{*}{$\Delta$} & \multirow{2}{*}{$t_{ss}$} & \multirow{2}{*}{$t_{pp}$} & \multirow{2}{*}{$t_{sp}$} & \multicolumn{2}{c|}{1D} & \multicolumn{2}{c|}{2D} \\
\cline{7-10} {} & {} & {} & {} & {} & {} & $E_{Edge}$ & Gap & $E_{Edge}$ & Gap \\
\hline
\hline
\textbf{Li} & 3.04 & 2.64 & 1.19 & 1.51 & 1.36 & \textbf{2.54} & 2.75 & \textbf{3.81} & 3.19 \\
\hline
\textbf{Be} & 2.15 & 3.73 & 2.25 & 2.23 & 2.35 & \textbf{0.00} & 5.23 & \textbf{2.61} & 6.36 \\
\hline
\textbf{Na} & 3.28 & 4.57 & 1.03 & 1.69 & 1.31 & \textbf{2.17} & 0.87 & \textbf{4.43} & 1.09 \\
\hline
\textbf{Mg} & 3.07 & 4.96 & 1.04 & 1.70 & 1.31 & \textbf{0.00} & 0.53 & \textbf{1.01} & 0.78 \\
\hline
\textbf{K} & 4.10 & 2.73 & 0.65 & 0.84 & 0.74 & \textbf{1.32} & 0.27 & \textbf{2.68} & 0.52 \\
\hline
\textbf{Ca} & 3.99 & 2.12 & 0.58 & 0.34 & 0.55 & - & 0.28 & \textbf{0.20} & 0.18 \\
\hline
\textbf{Rb} & 4.34 & 2.19 & 0.56 & 0.43 & 0.57 & - & 0.22 & \textbf{2.30} & 0.19 \\
\hline
\textbf{Sr} & 4.48 & 1.75 & 0.47 & 0.12 & 0.43 & - & 0.57 & - & 0.18 \\
\hline
\textbf{Cs} & 4.76 & 1.68 & 0.46 & 0.16 & 0.40 & - & 0.43 & - & 0.09 \\
\hline
\textbf{Ba} & 3.90 & 2.25 & 0.63 & 0.01 & 0.40 & - & 0.98 & - & 0.36 \\
\hline
\textbf{Fr} & 4.69 & 2.40 & 0.49 & 0.30 & 0.41 & - & 0.83 & - & 0.50 \\
\hline
\textbf{Ra} & 5.11 & 1.94 & 0.37 & 0.07 & 0.32 & - & 1.04 & - & 0.71 \\
\hline
\end{tabular}

\caption{Topological phase data of 1D chains and 2D honeycomb lattice alkali and alkaline earth materials. The results are based on the listed Hamiltonian parameters, which were obtained from fits to \textit{ab initio} calculations performed using FPLO, an atomic orbital-based DFT code \cite{fplo}. $E_{Edge}$ is the energy of the center of topological edge states in the gap with respect to the Fermi energy set to zero. All energies are given in eV, and the equilibrium lattice constants $a$ are in angstrom. DFT band structures and geometry relaxation data are shown in the Supplementary Material.}

\label{table:phase_table}

\end{table}

As is convenient for the topological analysis of two-band systems, we write the Hamiltonian \eqref{eq:hamiltonian} as $h(k)=d_0(k) \sigma_0+ {\bm d}(k)\cdot{\bm\sigma}$ in terms of a vector ${\bm\sigma}=(\sigma_x,\sigma_y,\sigma_z)$ of Pauli matrices acting in the orbital $(s,p)$ basis and the identity matrix $\sigma_0$.  Here $d_0(k)= -2t_-\cos(ak)$, 
\begin{equation}\label{eq:pauli_hamiltonian}
{\bm d}(k) = \left( 0,\ 2t_{sp}\sin(ak),\ 
 -2t_+\cos(ak) - {\Delta}/2\right),
\end{equation}
and $t_\pm=(t_{pp} \pm t_{ss})/2$. For generic values of parameters $h(k)$ respects spinless time-reversal symmetry $\cT:\  h(k)^*=h(-k)$ with $\cT^2=1$, which places it in class AI of the Altland-Zirnbauer classification \cite{AZ_classification_topological}. In space dimensions 1 and 2 this class allows only trivial topology and hence we do not expect any strong topological phases in our model. If, however, we include spatial symmetries, then weak topological phases become possible. The Hamiltonian \eqref{eq:hamiltonian} is inversion-symmetric, respecting $\cI:\ \sigma_z h(k)\sigma_z=h(-k)$. The two symmetries, $\cT$ and $\cI$, together enforce the absence of the $\sigma_x$ term in $h(k)$, as also manifest in Eq. \eqref{eq:pauli_hamiltonian}. This form of Hamiltonian implies a quantized winding number, 
\begin{equation}\label{winding}
    \gamma_C=\frac{1}{2\pi i}\int_{-\pi/a}^{\pi/a} dk\langle u(k)|\partial_k|u(k)\rangle,
\end{equation}
where $u(k)$ denotes the periodic part of the Bloch wavefunction of the occupied band. For a $2\times 2$ Bloch Hamiltonian $4\pi\gamma_C$ equals the solid angle swept out by ${\bm d}(k)$ as $k$ traverses the Brillouin zone. When ${\bm d}(k)$ is confined to a plane as is enforced here by the presence of the inversion symmetry, only $\gamma_C=0,\pm 1/2$ are possible which respectively correspond to a trivial and topological phase. The Hamiltonian \eqref{eq:hamiltonian} thus possesses a $Z_2$ topological invariant and is topologically equivalent to the SSH model \cite{ssh, ssh_soliton_observation, spin_charge_separation_ssh_topological}.

The main physical aspect distinguishing the topological from the trivial regime is that the band structure in the topological phase is highly $s$-$p$ hybridized, while the trivial phase is reached if the energy difference between the $s$ and $p$ levels is high enough such that no avoided crossing occurs between the upper and the lower band. This contrasts these systems from the dimerization-driven SSH model. However, this essential behavior would be absent if not specifically for the even-odd nature of that hybridization, which can be visualized in Fig. \ref{fig:Chain_Be_Ra_Results} (a) as $s$-$p$ hopping to the left and to the right of a given site having opposite signs.  If, for example, we replaced the $p$ orbital by another of the same energy but with even parity, the off-diagonal terms in the Hamiltonian \eqref{eq:hamiltonian} would not be of the form $\pm i\sin(ak)$  thus removing the possibility of a non-zero winding number around the origin in the Pauli matrix space. The Kitaev chain Hamiltonian has exactly the same off-diagonal element structure (in that case due to the form of the $p$-wave superconducting order parameter). Systems involving higher angular momentum states than $s$ or $p$ can also have even-odd mixing symmetry, similarly making them potential candidates for extending the reach of this mechanism to other classes of materials. Interesting candidates would be ones involving $p$ and $d$ bands, or $f$ and $d$ bands \cite{f_state_magnetic_topology} which often also involve strong correlations and magnetic phases \cite{zr_singlet}.

\begin{figure}
\includegraphics[width=\columnwidth]{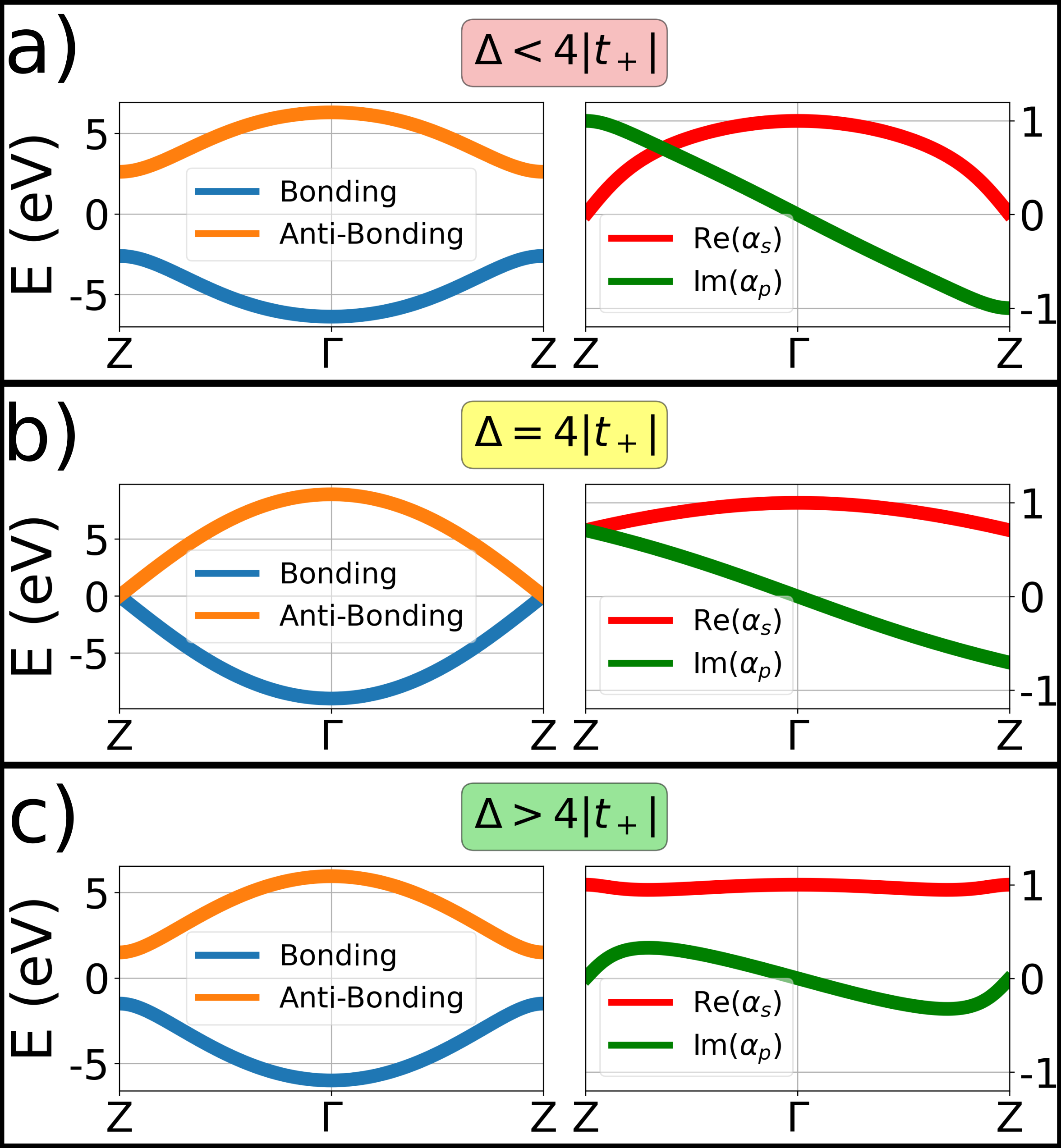}
\caption{1D chain band structure (left column) and the eigenstates of the lower band with respect to $k$ (right column) in different regimes; a) for the topological phase, b) at the phase transition point and c) in the trivial regime. The Fermi energy is set at $E=0$, and $t_{sp} > 0$ for all cases.}\label{fig:Chain_Topo_Phases}
\end{figure}

For systems with inversion symmetry and a $Z_2$ topological classification, it is possible to calculate the invariant by analyzing the inversion properties of the eigenstate representing the occupied fermionic band, in our case the lower, mainly $s$ band. We thus express the $k$-dependent wavefunction $\Psi\left(k\right)$ associated with this band as
\begin{equation}\label{eq:wavefunction}
\Psi(k) = \alpha_s(k)\phi_s + \alpha_p(k)\phi_p,
\end{equation}
where $\phi_s$ and $\phi_p$ are the wavefunctions associated with the $s$ and $p$ atomic states respectively, and $\alpha_s\left(k\right)$ and $\alpha_p\left(k\right)$ are normalized complex coefficients. With a suitable choice for the overall phase, we can write this wavefunction such that $\alpha_s\left(k\right)$ is purely real and $\alpha_p\left(k\right)$ is purely imaginary for all $k$ \cite{derriche_peierls}. Then, if we consider the eigenvalue $\gamma\left(k\right)$ of the inversion operator $\cI$ associated with $\Psi\left(k\right)$ at the two special $k$-points that map to themselves under inversion, namely $k=0$ and $k=\frac{\pi}{a}$, we can define a $Z_2$ invariant $\nu$ as
\begin{equation}\label{eq:invariant}
\nu = \gamma(0) \gamma\left(\frac{\pi}{a}\right).
\end{equation}
In this case, $\gamma\left(k\right)$ can only be equal to $1$ or $-1$. Thus, $\nu$ also either has the value 1  (trivial phase, $\Delta>4\left|t_+\right|$) or $-1$ (topological phase, $\Delta < 4\left|t_+\right|$ and $t_{sp} > 0$) \cite{inversion_sym, inversion_insulators}. This is consistent with the results displayed in Fig. \ref{fig:Chain_Topo_Phases}; for the topological phase, $\gamma\left(0\right)\ = 1$ and $\gamma\left(\frac{\pi}{a}\right) = -1$, while they are both equal to 1 for the trivial phase. At exactly the phase transition point $\Delta=4\left|t_+\right|$ the winding number is undefined since the system becomes gapless and it is not possible to uniquely identify the eigenstate belonging to the lower band. It is crucial to note here that the band structures in (a) and (c) are nearly identical in shape and structure. Consequently, looking solely at DFT band structures, one may not recognize that a system is in a topological phase; information about the electronic wavefunctions is necessary. Seeking only band crossings with Dirac-like dispersion and spin-orbit degeneracy breaking, the type of topology highlighted here can easily be missed. 

Importantly, the topological condition we have outlined only necessitates a non-zero $t_{sp}$ value; this guarantees avoided crossing between the bonding and antibonding bands occurs if the relationship between $\Delta$, $t_{ss}$ and $t_{pp}$ allows it, opening a gap that hosts a pair of topologically-protected edges states. It is true nevertheless that a very small $t_{sp}$ value can lead to a narrow gap, making the experimental detection of in-gap topological states harder. Thankfully however, as seen by the evolution of the hopping values in Table \ref{table:phase_table}, the magnitude of $t_{sp}$ follows the general trend of $t_{ss}$ and $t_{pp}$ since these hopping integrals depend on the orbital overlap of $ns$ and $np$ orbitals. Thus, materials with the right low-energy odd orbital phase relationships that exhibit significant same-orbital hopping as indicated by large band widths (obtained from DFT calculations for example) will tend to also have significant $t_{sp}$ magnitudes, making them good candidates to investigate this model's topological behavior if $\Delta$ allows crossing in the first place. The most obvious materials respecting this description are ones whose isolated constituent atoms have occupied $s$ states and unoccupied $p$ states near the Fermi energy that form crossing (or anticrossing with $t_{sp} \ne 0$) bands when brought together to make a solid. Other than the alkali and alkaline earth elements, this could potentially include transition metals with appropriate electronic configurations like Zn, Cd or Hg with filled $ns$ orbitals as their highest energy occupied states. 

\begin{figure}
\includegraphics[width=\columnwidth]{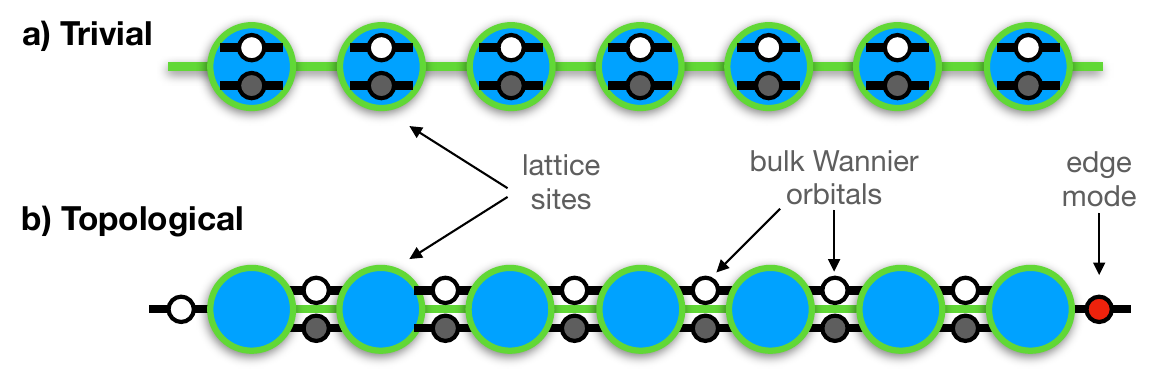}
\caption{Physical origin of the fractionally quantized polarization in the $s$-$p$ chain with $N$ sites. a) The trivial phase ($\gamma_C=0$), where $N$ Wannier orbitals are centered at the lattice sites. b) The topological phase ($\gamma_C=\pm \frac{1}{2}$), where $N-1$ Wannier centers are located midway between the lattice sites, forcing an electron to occupy one of two edge modes resulting in polarization $P=\pm \frac{e}{2}$.}\label{fig:Wannier_Diagram}
\end{figure}

In the special case $t_{pp}=t_{ss}$, the term in $h(k)$ proportional to $\sigma_0$ vanishes and the Hamiltonian acquires particle-hole symmetry, $\cP:\ \sigma_x h(k)^*\sigma_x=-h(-k)$ with $\cP^2=1$. This places the system in class BDI which admits integer topological classification in 1D \cite{10_fold_way}. The Hamiltonian then becomes topologically equivalent to the Kitaev chain where the topological invariant counts the number of Majorana zero modes localized at one end of the chain \cite{Kitaev, kitaev_exact, majorana_superconductors_review}. In our case the Hamiltonian acts in the space of $s$ and $p$ orbitals (and not in Nambu space) but zero modes still form when in the topological phase. The fact that the Be chains almost exhibit this symmetry since $t_{ss} \approx t_{pp}$ as seen in Table \ref{table:phase_table} highly increases the potential for experimental observation of its topological edge states since their centering in the large band gap shown in Fig. \ref{fig:Chain_Be_Ra_Results} (c) makes it harder for perturbations to shift their energy up or down which would lead to them being swallowed by the bulk bands. Like in the SSH model, the physical observable associated with the topological phase are the two fractional polarization states $P=\pm \frac{e}{2}$ which can be interpreted as the last valence electron residing in one of the two degenerate end modes \cite{geometric_polarization_topological_phase}. An intuitive picture for this mechanism is given in Fig. \ref{fig:Wannier_Diagram}, which shows the emergence of edge Tamm states in the topological phase \cite{Tamm_original, tamm_polaron}.

\begin{figure}
\includegraphics[width=\columnwidth]{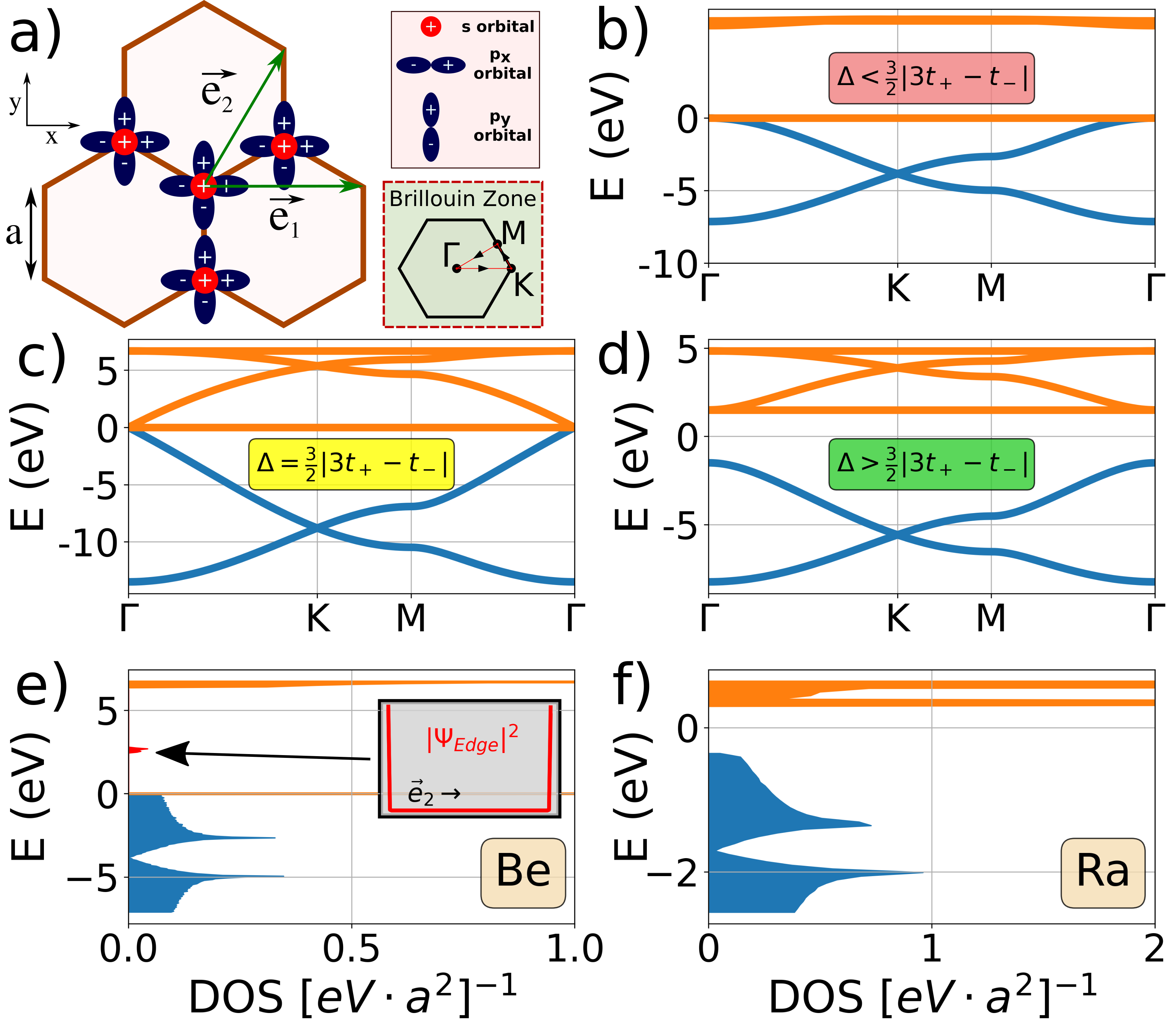}
\caption{Results for a 2D $s$-$p$ honeycomb lattice. a) Honeycomb lattice with an $s$, $p_x$ and $p_y$ orbital at each site. $\va{e}_1$ and $\va{e}_2$ are the primitive lattice vectors, and high symmetry points in the associated Brillouin Zone are shown. b) Band structure in the topological phase, c) at the phase transition point and d) in the trivial phase. e) Density of states of a honeycomb strip that is infinite along the $\va{e}_1$ direction with a thickness of 100 unit cells in the $\vec{e}_2$ direction made of Be atoms, and f) of Ra atoms. The inset shows the momentum-averaged combined charge density of the two gap topologically protected dispersive edge modes. The Fermi energy is set at $E=0$.}\label{fig:Honeycomb_Be_Ra_Results}
\end{figure}

While physically enlightening, one-dimensional models are often difficult to realize experimentally although we here envision the possibility of forming 1D chains at step edges of a substrate which is inert regarding covalent interactions with the chain atoms \cite{nanowires_mos2_on_steps_substrate}. We thus extend our analysis by considering a 2D honeycomb lattice governed by the same even-odd interatomic hybridization and demonstrate that the conclusions we reached for the 1D system also apply in higher dimensions. The honeycomb lattice was specifically chosen because, in analogy with graphene, its energy spectrum is gaped for all $\va{k}$ away from two Dirac points and can be easily turned insulating. This is unlike most square lattice systems which typically exhibit large Fermi surfaces, distinguishing this system from the BHZ model \cite{quantum_spin_hall_effect_original,colloq_topo}. It must be made clear that the chains and honeycomb systems with the DFT-relaxed lattice constants listed in Table \ref{table:phase_table}, especially the ones composed of heavier elements, may not all be fully stable experimentally but they are useful to demonstrate the strong potential of the charge-based even-odd hybridization-induced topology this manuscript is centered on and to showcase the atomic number-dependence of those topological properties. As shown in Fig. \ref{fig:Honeycomb_Be_Ra_Results} (a), each site of the honeycomb lattice in the $x$-$y$ plane hosts an $s$, $p_x$ and $p_y$ orbital, and nearest neighbor hopping between all of those is allowed. Because of the lower number of valence electrons per atom in these systems compared to carbon's four, the highest-energy occupied states are $\sigma$-bonded $sp_2$ orbitals, so the $p_z$ $\pi$-bonding here is not active without doping contrary to graphene for which they are the principal active electronic states. The lowest-energy empty bands in graphene correspond to those $p_z$ $\pi$-bonding states and they do not hybridize with the in-plane $\sigma$-bonded orbitals, making spin-orbit coupling necessary for a band gap to open while the alkali and alkaline earth systems only need even-odd hybridization. The resulting 6-band Hamiltonian is most easily written in the chiral basis $(s_A,p_{A+},p_{A-};s_B,p_{B+},p_{B-})$  where $s_\eta$ denotes the $s$ orbital on $\eta=A,B$ sublattice and $p_{\eta\pm}=p_{\eta x}\pm i p_{\eta y}$. It reads  
\begin{equation}\label{eq:h2D}
h(\va{k}) = \begin{pmatrix}
h_0 & h_{AB}(\va{k}) \\
h_{AB}(\va{k})^\dag & h_0
\end{pmatrix},
\end{equation}
where $h_0={\rm diag}(-\Delta,\Delta,\Delta)/2$, 
\begin{equation}\label{eq:hAB}
h_{AB}(\va{k}) = \begin{pmatrix}
t_{ss} z_0(\va{k}) & t_{sp} z_{2}(\va{k}) &  t_{sp} z_{1}(\va{k}) \\
t_{sp} z_{1}(\va{k}) & t_{pp} z_0(\va{k}) &  t_{sp} z_{1}(\va{k}) \\
t_{sp} z_{2}(\va{k}) & t_{sp} z_{2}(\va{k}) & t_{pp} z_0(\va{k})
\end{pmatrix},
\end{equation}
and $z_\mu(\va{k})=e^{i\va{k}\cdot\va{a}_1}+e^{i(\va{k}\cdot\va{a}_2-\mu \frac{2\pi}{3})}+e^{i(\va{k}\cdot \va{a}_3+\mu \frac{2\pi}{3})}$. Here $\va{a}_1=a\vu{y}$, $\va{a}_{2,3}=a(-\sqrt{3}\vu{x}\pm\vu{y})/2$ are three nearest neighbor vectors in the honeycomb lattice \cite{semenoff_graphene_tight_binding, graphene_tb_model}.

Using again the parameters in Table \ref{table:phase_table} (with the $p_x$ and $p_y$ atomic levels being degenerate), we compare the band structure in Figs. \ref{fig:Honeycomb_Be_Ra_Results} (b-d). Interestingly, along with lower energy mainly $s$ bands and higher energy mainly $p$ bands, two completely flat purely $p$ bands appear. We identify these as linear combinations of localized $p$ states in the honeycomb bulk discussed previously \cite{p_hex_system, semenoff_graphene_edge_magnetism}. It is interesting to note that the flatness of these bands is not perturbed by the introduction of $s$-$p$ mixing into the model. Such flat bands are appealing due to their propensity to form strongly correlated phases in the presence of interactions and have been studied in optical and electronic systems \cite{hex_electronic}. An interesting feature of the Hamiltonian described by equations \ref{eq:h2D} and \ref{eq:hAB} is that, similarly to the 1D chain, a band inversion that occurs when $\Delta = 3|t_{ss} + \frac{t_{pp}}{2}| = \frac{3}{2}|3t_+ - t_-|$. As illustrated in Fig. \ref{fig:Honeycomb_Be_Ra_Results} (b-d), at the phase transition three bands become degenerate at the $\Gamma$ point whereby the lower flat band detaches from $p$-bands and attaches to $s$-bands. Consequently, the Fermi energy for half-filled alkaline earth honeycomb systems in the topological phase lies right under this set of localized $p$ states, which have to be filled to access the topological end modes in the gap. By projecting the Hamiltonian \eqref{eq:h2D} onto the degenerate subspace it is straightforward to derive an effective 3-band $\va{k} \cdot \va{p}$ Hamiltonian for this transition,  
\begin{equation}\label{heff}
h_{\rm eff}(\va{k}) = \begin{pmatrix}
0 & 0 & -v k_- \\
0 & 0 & v k_+ \\
-v k_+ & v k_- & m \\
\end{pmatrix}.
\end{equation}
\noindent Here $k_\pm=k_x\pm ik_y$, $v=3t_{sp}a/2$ and $m=\Delta-\frac{3}{2}|3t_+ - t_-|$. The energy eigenvalues are $(m, \pm\sqrt{m^2+v^2k^2})$. At the transition point marked by $m=0$ the spectrum consists of a Dirac cone and a flat band, both centered at zero energy. Although once again there is in general no strong topology allowed in the Hamiltonian belonging to AI symmetry class in 2D, we can understand the emergence of the edge modes by regarding the 2D Bloch Hamiltonian $h(k_x,k_y)$ placed on a long strip as a collection of 1D Hamiltonians $h_{k_y}(k_x)$ parametrized by the transverse momentum $k_y$. We find that, when the system resides in the inverted regime, the flat band contributes a quantized winding number $\gamma_C$ for each transverse momentum $k_y$ and the invariant $\nu$ defined in Eq. \eqref{eq:invariant} is equal to $-1$. Like in the 1D chain case, this results in a localized end mode with the energy inside the gap for each $k_y$. These end modes then form a dispersive edge state visible in Fig. \ref{fig:Honeycomb_Be_Ra_Results} (e). Note that as $h_{k_y}(k_x)$ is not particle-hole symmetric we do not expect the dispersive edge modes to be centered in the gap.

\section{Conclusion}

These results open the way for experimental studies of an underexplored type of topological materials favoring lighter elements in contract to spin-orbit coupling driven topological systems; any material that exhibits the requisite $s$-$p$ hybridization in its low-energy sector such as the alkali and alkaline earth elements is a potential candidate for the observation of these properties. Notably, due to naturally hosting one more valence electron per atom than the alkali, alkaline earth materials are of particular interest since their Fermi energy in 1D chains lies directly at the topologically-protected edges modes, making their experimental detection significantly easier. Beryllium chains in particular stand out as the most promising even-odd hybridisation-based topological insulators because of their near particle-hole symmetry and large band gaps strongly protecting the end modes from perturbations compared to other candidates \cite{be_chain_dft_electronic_structure}. Beryllene, The 2D honeycomb Be system, has been experimentally realized \cite{beryllene} and would be an excellent system to explore the ramifications of our model, especially since Be-based materials such as Be$_3$X$_2$ (X = C, Si, Ge, Sn) \cite{be_candidate_chinese} have been predicted to be Dirac semimetals and to possess significant contributions from surfaces states to the density of states at the Fermi energy \cite{be_3d_surface_states, polysilyne}. Interestingly, while other theoretical studies of 2D Be hexagonal structures predict similar interatomic distances as the one used in this work \cite{beryllene_prediction, beryllene_superconducticity_topo}, a nearest neighbor interatomic distance of 1.3 \angstrom{} was reported for experimentally-realized beryllene; the fact that this distance is considerably smaller than 2.15 \angstrom{} actually raises its appeal as a candidate for the detection of charge-based topological edge states since an enhancement of bandwidths leads to larger anticrossing-induced band gaps.  Other low-dimensional candidates to explore $s$-$p$ topological physics which include alkaline earth elements are CaB$_6$ nanowires \cite{nanochain} or MgB$_2$ \cite{2d_example_mgbh}.

\section{Data Availability Statement}

Any data that support the findings of this study are included within the article (and any supplementary files).

\section{Acknowledgements}

This research was undertaken thanks in part to funding from the Max Planck-UBC-UTokyo Center for Quantum Materials and the Canada First Research Excellence Fund, Quantum Materials and Future Technologies Program, as well as by the Natural Sciences and Engineering Research Council (NSERC) for Canada.


%%%% Start of Supplementary Materials %%%
\include{supplementary}

\end{document}

%% file: supplementary.tex
%Changing the labeling of Figures for supplementary information, and other style things 
\graphicspath{{supp_figs/}}
\renewcommand{\thefigure}{S.\arabic{figure}}
\setcounter{figure}{0} 
\onecolumngrid
\pagenumbering{roman}

\section{Supplementary Material: DFT Structural Relaxation and Band Structure Data}

\begin{figure}[h]
\includegraphics[width=\columnwidth]{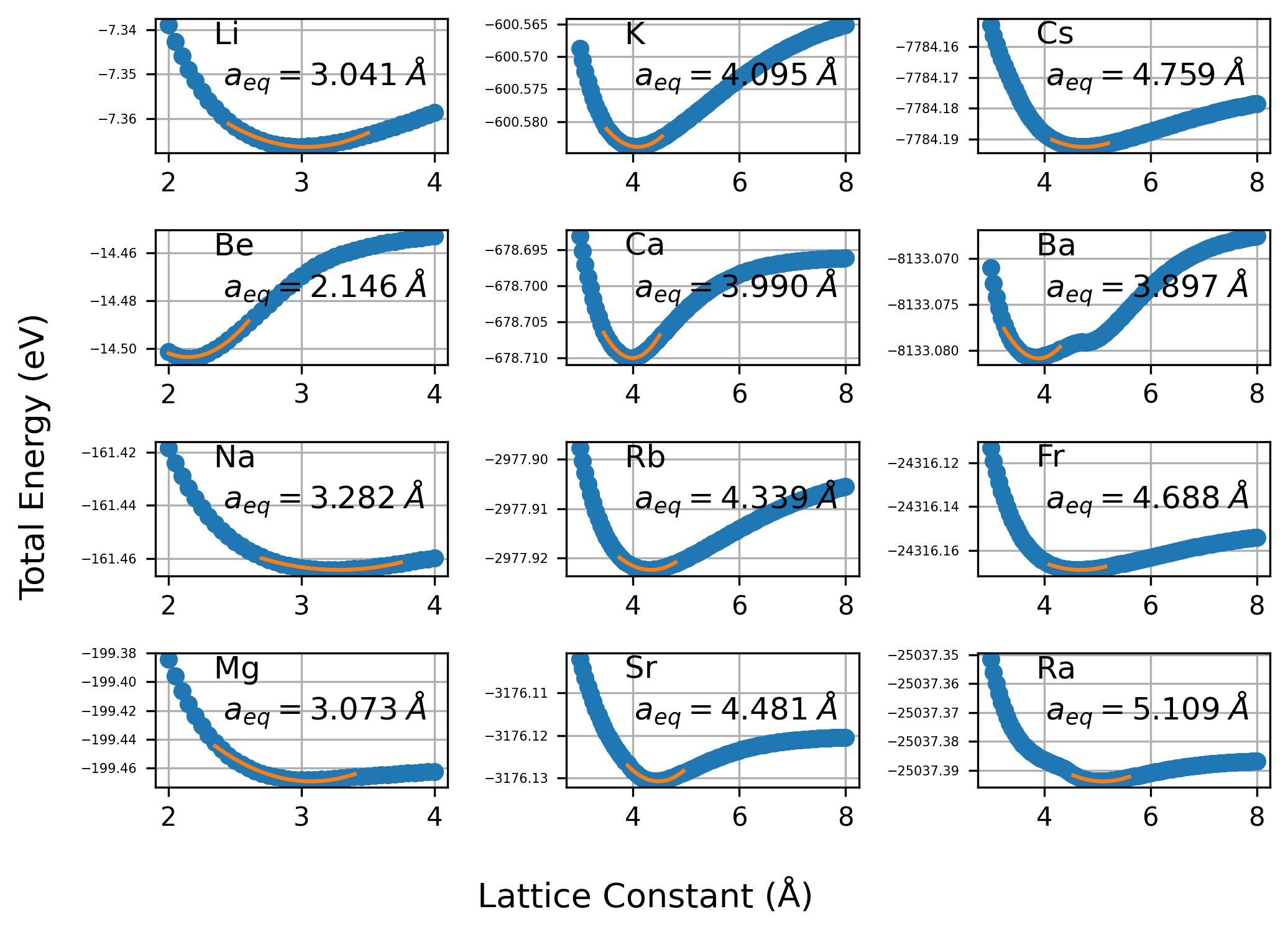}
\caption{Total energy with respect to the lattice constant for all alkali and alkaline earth chains calculated with DFT. The quadratic fit performed around the energy minima to extract the equilibrium lattice constants $a_{eq}$ is shown in orange for all elements.}\label{fig:1d_chain_equilibrium_constants}
\end{figure}

We present in this Supplementary Material the data for the DFT geometry relaxation process we have performed for the alkali and alkaline earth equally-spaced 1D chains. We have calculated the total energy of such chains for a large range of nearest neighbor distances to determine the equilibrium lattice constant for each element, as shown in Fig. \ref{fig:1d_chain_equilibrium_constants}. This was done through quadratic fits of the energy minimum basins to extract accurate lattice constants.

%%%%%
%%%%%%

\begin{figure}[h]
\includegraphics[width=\columnwidth]{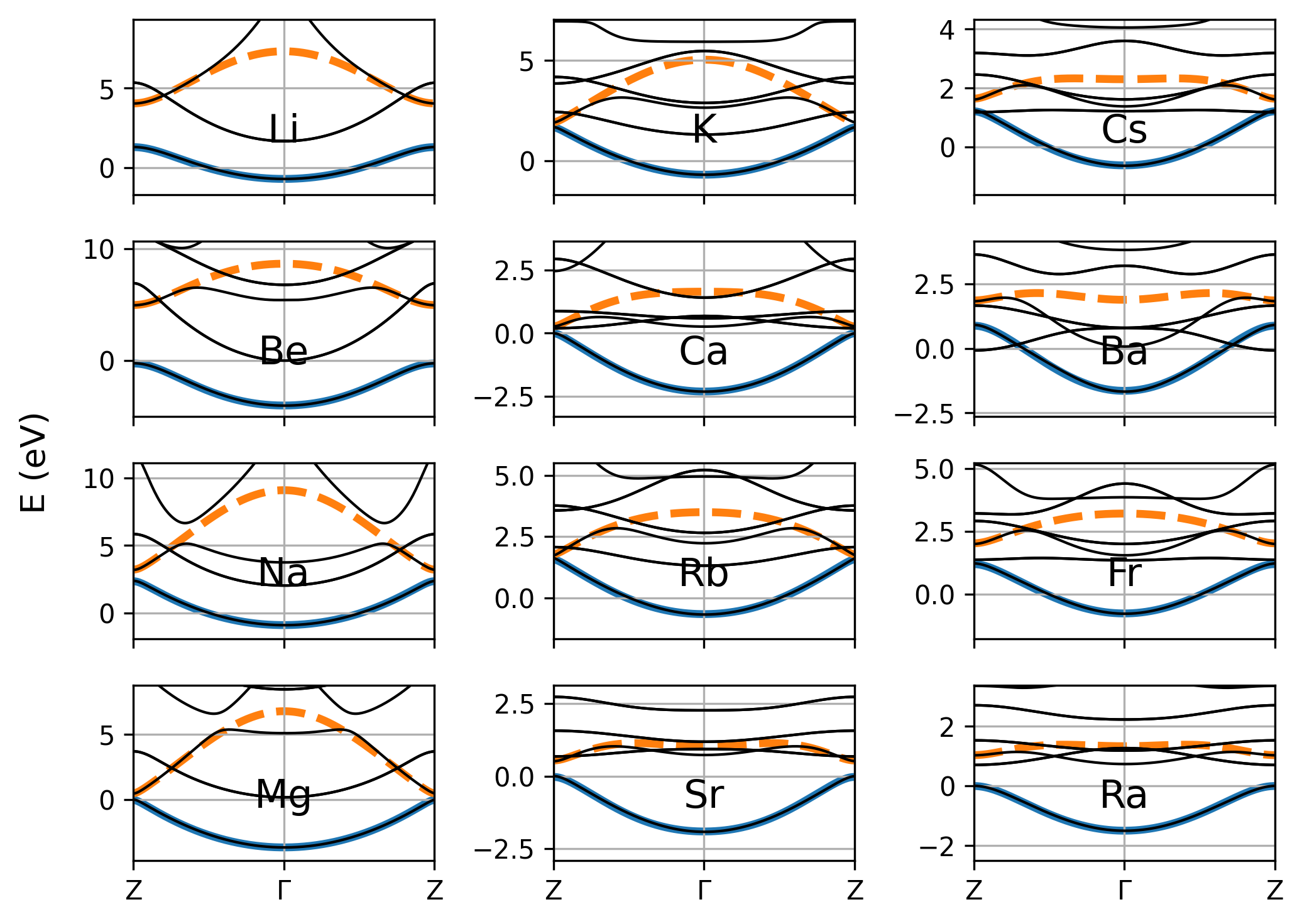}
\caption{Band structures of all alkali and alkaline earth relaxed 1D chains. The black lines are the DFT-calculated energy dispersions, while the blue and orange lines are respectively the bonding and antibonding bands corresponding to the tight binding fit we have performed on the $ns$ and $np$ parts of the appropriate DFT bands. The Fermi energy is at $E=0$ for all plots.}\label{fig:1d_chain_all_bands}
\end{figure}

\newpage
The band structures of the alkali and alkaline earth chains with those relaxed lattice constants were then calculated through DFT, and the mainly-$ns$ and mainly-$np$ bands near the Fermi energy were fitted using the two-band tight binding $s$-$p$ model detailed in the main text in order to extract the Hamiltonian parameters. This data, along with the results of fitting, is displayed in Fig. \ref{fig:1d_chain_all_bands}. It must be noted that the DFT band data here naturally contains more bands than our two-band system, such as the $p_x$ and $p_y$ $\pi$-bonding band in between the fitted bands, but these can ignored in this model since $\pi$-bonded valence orbitals in an actual realization of such chains should be significantly increased in energy.

%%%%%
%%%%%%

\begin{figure}[h]
\includegraphics[width=\columnwidth]{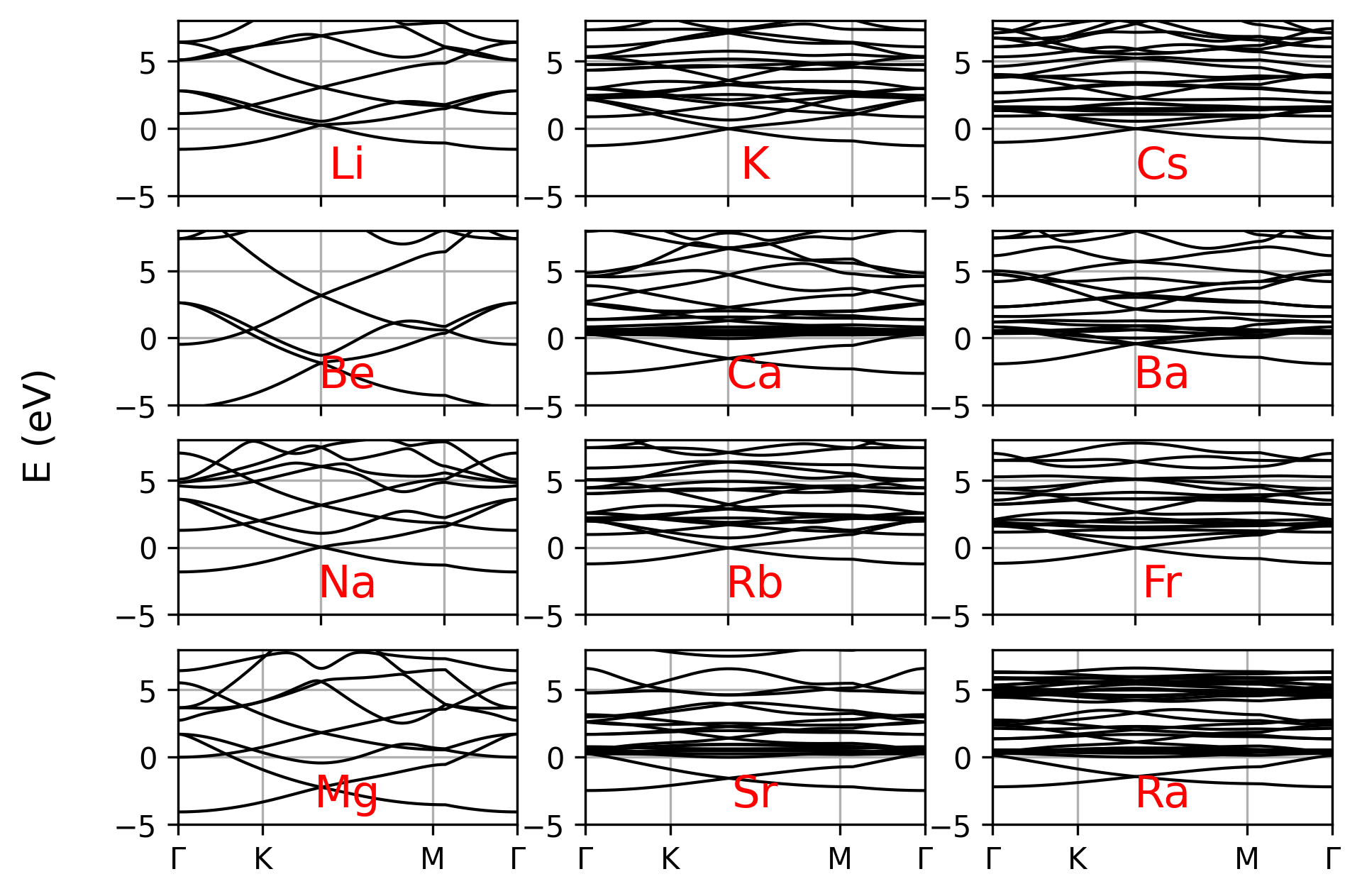}
\caption{DFT-Calculated band structures of all alkali and alkaline earth 2D honeycomb structures. The Fermi energy is at $E=0$ for all plots.}\label{fig:2d_honeycomb_all_bands}
\end{figure}

\newpage
Furthermore, we show the band structures of alkali and alkaline earth 2D honeycomb structures computed with DFT by using the equilibrium nearest neighbor distances obtained for 1D chains in Fig. \ref{fig:2d_honeycomb_all_bands}. Like for the 1D DFT band structures, other bands which do not enter in our model are present, notably $p_z$ $\pi$-bonding bands which are present in the gap between the $s$ and $p$ $\sigma$-bonded bands.

%%%%%
%%%%%%